\documentclass[9pt]{article}
\usepackage{graphicx}

\renewcommand{\in}{\raise -3pt\hbox{\scriptsize in}}
\newcommand{\out}{\raise -3pt\hbox{\scriptsize out}}

\begin{document}

\begin{flushright}
CPHT--RR 059.0904 \\
LPT--04.099  
\end{flushright}

\vspace{2cm}
\vspace{\baselineskip}

\begin{center}
\textbf{\LARGE Exotic hybrid mesons in hard electroproduction} \\
\vspace{2\baselineskip}
{\large I.V.~Anikin$^{a,d}$,\, B.~Pire$^b$,\,
 L.~Szymanowski$^{c,e}$,\, O.V.~Teryaev$^{a}$,\, S.~Wallon$^d$
}
\\
\vspace{2\baselineskip}
{\it ${}^a$\,Bogoliubov Laboratory of Theoretical Physics, JINR, 141980 Dubna,
Russia \\[0.5\baselineskip]
${}^b$\,CPHT \footnote{Unit{\'e} mixte 7644 du CNRS}, {\'E}cole
Polytechnique, 91128 Palaiseau, France \\[0.5\baselineskip]
${}^c$\,Soltan Institute for Nuclear Studies, Warsaw, Poland \\[0.5\baselineskip]
${}^d$\,LPT \footnote{Unit{\'e} mixte 8627 du CNRS}, Universit{\'e} Paris-Sud, 91405-Orsay,
France
 \\[0.5\baselineskip]
${}^e$\,Phys. Th\'eor. Fondam., Inst. de Physique,
Univ. de Li{\`e}ge, B-4000 Li{\`e}ge, Belgium
\\
}
\vspace{2\baselineskip}
\textbf{Abstract}\\
\vspace{1\baselineskip}
\parbox{0.9\textwidth}
{We estimate the sizeable cross section for deep exclusive electroproduction of an exotic
$J^{PC}=1^{-+}$ hybrid meson in the Bjorken regime.  The production amplitude scales like
the one for usual meson electroproduction, {\it i.e.} as
$1/Q^2$. This is due to the non-vanishing  leading twist distribution
amplitude for the hybrid meson, which may be normalized thanks to its relation to the
energy momentum tensor and to the QCD sum rules technique. The hard amplitude is
considered up to next--to--leading order in $\alpha_{S}$ and we explore the consequences
of fixing the renormalization scale ambiguity through the BLM procedure.
We study the particular case where the hybrid meson decays through a $\pi\eta $ meson pair.
We discuss the $\pi\eta$ generalized distribution amplitude and then calculate
the production amplitude for this process.
We propose a forward-backward asymmetry in the production of $\pi$ and $\eta$ mesons
as a signal for the hybrid meson production.
We briefly comment on hybrid electroproduction at very high energy, in
the diffractive limit where a QCD Odderon exchange mechanism should
dominate. The conclusion of our study is that hard electroproduction is
a promissing way to study exotic hybrid mesons, in particular at
JLAB, HERA (HERMES) or CERN (Compass).}
\end{center}

\newpage
\section{ Introduction}
\vspace{.5cm}

\noindent
Within quantum chromodynamics, hadrons are described in terms of quarks, anti-quarks and
gluons. The usual, well-known, mesons are supposed to contain quarks and
anti-quarks as valence
\footnote{The valence degrees of freedom define the charge and other quantum numbers of
corresponding hadrons, while the sea configurations do not change the quantum numbers.}
degrees of freedom while gluons play the role of carrier of interaction,
{\it i.e.} they remain hidden in a background.
On the other hand, QCD does not prohibit the existence of the explicit gluonic degree of freedom
in the form of a vibrating flux tube, for instance. The states where the $q\bar q g$ and
$gg$ configurations are dominating, hybrids and glueballs,
are of fundamental importance to understand the dynamics of quark confinement and the
nonperturbative sector of quantum chromodynamics \cite{Close}--\cite{Bernard}.

\noindent
The study of these hadrons outside the constituent quark models, namely exotic hybrids, is the main 
reason of the present paper. We investigate how hybrid mesons with $J^{PC}=1^{-+}$
may be studied through the so-called hard reactions.
We focus on deep exclusive meson electroproduction
(see, for instance \cite{Goe}) which is well described in the framework of the collinear
approximation where generalized parton distributions (GPDs) \cite{Muller:1994fv} and distribution
amplitudes\cite{ERBL} describe the nonperturbative parts of a factorized amplitude 
\cite{Collins:1997fb}.

In a previous paper \cite{APSTW1} we showed that contrarily to naive expectations, the
amplitude for the electroproduction of an isotriplet exotic meson with $J^{PC} = 1^{-+}$ may be
written in a very similar way as the amplitude for non-exotic vector meson electroproduction.
The main observation of our work was that the
quark-antiquark correlator on the light cone includes a gluonic component due to gauge
invariance and leads to a leading twist hybrid light-cone distribution amplitude.
In this paper, we extend our analysis of the electroproduction process and
calculate the differential cross section as a function of $Q^2$.

\noindent
We also study the hybrid meson  as a resonance in the reaction $e\,p\to e\,p\,(\pi^0\eta)$.
The first experimental investigation of the hybrid with $J^{PC}=1^{-+}$ as the resonance in
$\pi^-\eta$ mode was implemented by the Brookhaven collaboration E852 \cite{E852}.
Present candidates for the hybrid states with $J^{PC}=1^{-+}$ include $\pi_{1} (1400)$ which is
mostly seen through its $\pi\eta$ decay and $\pi_{1} (1600)$ which is seen through its
$\pi \eta' $ and $\pi \rho $ decays \cite{RPP}.
Theoretically these states are objects of intense studies \cite{Close}, mostly through
lattice simulations \cite{Bernard}.

\vspace{.5cm}
\section{Hybrid meson production amplitude}
\vspace{.5cm}

\noindent
We propose to study the exotic hybrid meson by means of its deep exclusive
electroproduction, {\it i.e.}
\begin{eqnarray}
\label{pr}
e(k_1)\, + \, N(p_1)\,\to\,e(k_2)\,+ H(p)\,+\,N(p_2),
\end{eqnarray}
where we will concentrate on the subprocess:
\begin{eqnarray}
\label{pr2}
\gamma^*_L(q)\, + \, N(p_1)\,\to\, H_L(p)\,+\,N(p_2)
\end{eqnarray}
when the baryon is scattered at small angle.
This process is a hard exclusive reaction due to the transferred momentum $Q^2$
is  large ( Bjorken regime).
Within this regime, a factorization theorem is valid,  at
the leading twist level, which claims that a partonic subprocess part described in perturbative
QCD (pQCD)
can be detached from universal soft parts, which are generalized parton distributions and meson
distribution amplitudes.
Below we will analyze in more details how this factorization theorem applies to the  process
under study.

\noindent
In this paper, the main object of our investigation is the isotriplet of mesons with quantum 
numbers
$J^{PC}=1^{-+}$. Such mesons can be named as exotic mesons because they do not exist within
the usual quark model. To illustrate the latter we shortly remind the main steps of
the description and classification of meson states in the quark model.

\vspace{.5cm}
\subsection{Quark model and spectroscopy}
\vspace{.5cm}

\noindent
It is well known that in the quark model the hadrons, mesons and baryons,
are  bound states of quark-antiquark or three-quarks systems.
Let us consider the mesons, {\it i.e.} the quark-antiquark systems.
Their  total angular momentum results  from the summation of spin
$S$ and orbital $L$ angular momenta of quarks. Neglecting a spin-orbital
interaction, the quantum numbers $S$ and $L$ may be considered
as additional quantum numbers for the classification of hadron states.
Therefore, the eigenvalues of the squares of the angular momenta read:
\begin{eqnarray}
\label{mom}
&&{\bf J}^2=J(J+1)\quad {\bf S}^2=S(S+1) \quad {\bf L}^2=L(L+1),
\nonumber\\
&& {\bf J}= {\bf S} + {\bf L},
\end{eqnarray}
where the number $L$ may take all positive integer values (including zero).
The meson octets correspond to the case where $S= 0,\,1$. For given values
of $S$ and $L$, the total angular momentum $J$ can take the values
\begin{eqnarray}
\label{Jval}
J\,=\,S+L,\,S+L-1,\, ...\,, \left| S-L \right|.
\end{eqnarray}
The values $S$ and $L$ are related to
the $C$- and $P$-parity of the quark-antiquark system in the form:
\begin{eqnarray}
\label{CP}
C=(-)^{L+S}, \quad P=(-)^{L+1}.
\end{eqnarray}
Consequently, in the quark model, the quantum numbers $S$, $L$, $J$,
$P$, $C$ and the relations between them (\ref{Jval}), (\ref{CP})
allow to introduce the following classification of the meson states~:
\begin{itemize}
\item $S=0,\, L=J $~:
\begin{eqnarray}
\label{S0}
J^{PC}= 0^{-+}, \, 1^{+-},\,2^{-+},\,3^{+-},\, ...
\end{eqnarray}
\item $S=1,\, L=0,\, J=1$~:
\begin{eqnarray}
\label{S1L0}
J^{PC}= 1^{--}
\end{eqnarray}
\item $S=1,\, L=1,\, J=2,\,1,\,0$~:
\begin{eqnarray}
\label{S1L1}
J^{PC}= 0^{++}, \, 1^{++},\,2^{++}
\end{eqnarray}
\item $S=1,\, L=2,\, J=3,\,2,\,1$~:
\begin{eqnarray}
\label{S1L2}
J^{PC}= 1^{--}, \, 2^{--},\,3^{--}
\end{eqnarray}
\end{itemize}
and so on.
From this, one can see the mesons with
 $J^{PC}= 0^{--}, \, 0^{+-},\, 1^{-+},\, ...\,,$
are forbidden. However, such mesons may be described beyond the quark model.
Indeed, we may add an extra degree of freedom (a gluon, for instance)
to get the needed quantum numbers, see for instance \cite{Jaffe}. 
Below we will consider this case in more details.

\vspace{.5cm}
\subsection{Kinematics and leading order amplitude}
\vspace{.5cm}

\begin{figure}
$$\includegraphics[width=12cm]{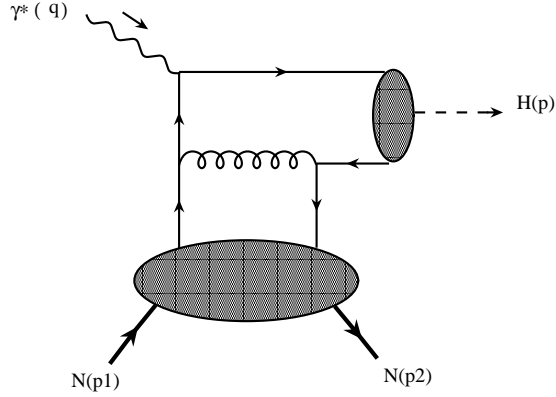}$$
\caption{Typical diagram describing the electroproduction of a
meson at lowest order. The grey blobs are non-perturbative matrix elements, namely
the meson distribution amplitude and the nucleon generalized parton distribution }
\label{diagram}
\end{figure}

\noindent
Let us fix the kinematics of the deep electroproduction process. We are interested in the
scaling regime where the virtuality of the
photon $Q^2=-q^2$ is large and scales with the energy of the process.
We denote by $p_1$ ($p_2$) the momentum of the incoming (outgoing) nucleon,
while $p$ is the momentum of the longitudinally polarized hybrid meson of mass $M_{H}$.
We construct the average momentum
$\overline{P}$ and transferred momentum $\Delta$:
\begin{eqnarray}
\label{PDel}
&&\overline{P}=\frac{p_2+p_1}{2}, \quad \Delta=p_2-p_1, \quad \Delta^2=t.
\end{eqnarray}
It is convenient to introduce the following two light-cone vectors:
\begin{eqnarray}
\label{nnst}
n^*=\Lambda (1,{\bf 0}^T,1), \quad n=\frac{1}{2\Lambda} (1,{\bf 0}^T,-1), \quad
n^*\cdot n=1,
\end{eqnarray}
where $\Lambda$ is an arbitrary dimensionful parameter
\footnote{Within the infinite momentum frame, the parameter $\Lambda$ may be
chosen as $\overline{P}^+$. Here, $A^+=(A^0+A^3)/\sqrt{2}$.}.
Then the Sudakov decompositions for  all the
relevant momenta take the form:
\begin{eqnarray}
\label{Sd}
&&\Delta_\mu=-2\xi n^*_\mu +\xi \overline{M}^2 n_\mu + \Delta^T_\mu,
\quad \Delta^T\cdot n=\Delta^T\cdot n^*=0,
\nonumber\\
&&\overline{P}_\mu=n^*_\mu+\frac{\overline{M}^2}{2}n_\mu , \quad \overline{P}^2=
\overline{M}^2,
\quad \xi\leq \frac{\sqrt{-\Delta^2}}{2\overline{M}}\leq 1,
\nonumber\\
&&q_\mu=-2\tilde\xi n^*_\mu +\frac{Q^2}{4\tilde\xi}n_\mu, 
\nonumber\\
&&p_\mu=q_\mu-\Delta_\mu=2(\xi-\tilde\xi) n^*_\mu+
\biggl( \frac{Q^2}{4\tilde\xi}-\xi\overline{M}^2 \biggr)n_\mu -\Delta^T_\mu.
\end{eqnarray}
Here, the parameters $\xi$ and $\tilde\xi$ are related by
\begin{eqnarray}
M_{H}^2=4(\xi-\tilde\xi)
\biggl(\frac{Q^2}{4\tilde\xi}-\xi\overline{M}^2 \biggr)+\Delta_T^2.
\end{eqnarray}
The photon longitudinal polarization vector can be written as
\begin{eqnarray}
\label{phpol}
\varepsilon_{L\,\mu}=\frac{2\tilde\xi}{Q}n^*_\mu+\frac{Q}{4\tilde\xi} n_\mu,
\end{eqnarray}
where the notation $Q=\sqrt{Q^2}$ is introduced.

\noindent
The leading order amplitude for the process (\ref{pr})
corresponding to the diagrams in Fig. \ref{diagram} is
\begin{eqnarray}
\label{amp01}
{\cal A}=\varepsilon_L^\mu \int d\eta\,e^{iq\cdot\eta}
\langle N(p_2)\, H(p) |
\frac{\delta S}{\delta A_\mu(\eta)} |N(p_1)\rangle ,
\end{eqnarray}
where the $S$-matrix is given by
\begin{eqnarray}
S=T{\rm exp}\biggl\{ i\int d^4 x \biggl({\cal L}_{QCD}(x)+{\cal L}_{QED}(x)
\biggr)\biggr\}.
\end{eqnarray}
Applying the factorization theorem, this amplitude can be written 
at leading twist and when $-t\ll Q^2$ as
\begin{eqnarray}
\label{AmF}
{\cal A}=\int\limits_{0}^{1} dz \int\limits_{-1}^{1} dx\,
\Phi_H(z,\mu^2,\mu_R^2) \, H(x,z,Q^2,\mu^2,\mu^2_R) \,
F(x,\mu^2,\mu_R^2)\equiv
\Phi_H\, \otimes \,H \, \otimes\, F,
\end{eqnarray}
where the parameters $\mu^2$ and $\mu^2_R$ are the factorization and
renormalization scales, respectively. Throughout  this paper, we will adopt the convention that $\mu=\mu_R$
\footnote{ The arguments in favour of such a choice in the case of the pion form factor are
discussed {\it e.g.} in \cite{RadE, Melic}}.
In Eq.(\ref{AmF}), $H$ is the hard part of amplitude which is controlled by  perturbative QCD.
The hybrid meson distribution amplitude $\Phi_H$ describes the transition from  the partons
 to the meson, and $F$ denotes generalized parton distributions which are related to
nonperturbative matrix elements of bilocal operators between different hadronic states.

\noindent
More precisely, the factorized amplitude (\ref{AmF}) may be written as :
\begin{eqnarray}
\label{amp02}
{\cal A}_{(q)}= \frac{e\pi\alpha_s f_{H} C_F}{\sqrt{2}N_c Q}
\biggl[ e_u {\cal H}_{uu}^- -e_d {\cal H}_{dd}^-\biggr] {\cal V}^{(H,\,-)},
\end{eqnarray}
where
\begin{eqnarray}
\label{softin1}
{\cal H}_{ff}^\pm &=&
\int\limits_{-1}^{1}dx \biggl[
\overline{U}(p_2)\hat n U(p_1) H_{ff^{\prime}}(x) +
\overline{U}(p_2)\frac{i\sigma_{\mu\alpha} n^{\mu}\Delta^{\alpha}}{2M}U(p_1)
E_{ff^{\prime}}(x) \biggr]
\nonumber \\
&&
\biggl[
\frac{1}{x+\xi-i\epsilon}\pm\frac{1}{x-\xi+i\epsilon}\biggr],
\nonumber\\
\label{softin2}
{\cal V}^{(M,\,\pm)}&=&
\int\limits_{0}^{1} dy \phi^{M}(y)\biggl[
\frac{1}{y}\pm\frac{1}{1-y}
\biggr].
\end{eqnarray}
Here, functions $H$ and $E$ are standard leading twist GPD's and their properties are
fairly well-known (see, for instance, a review of Diehl in \cite{Muller:1994fv}).
In (\ref{softin1}), we include the definition of ${\cal H}_{ff}^+$ and 
${\cal V}^{(M,\,+)}$ which will be useful for the comparison with the $\rho$ meson 
case. The hybrid meson distribution amplitude which enters Eqn. (\ref{softin2}) is a new object
and we will carefully study it in the next subsection.
Note that the simple pole over $y$ in (\ref{amp02}) does not lead to
any infrared divergency if the function $\phi^{H}(y)$ vanishes
when the fraction $y$ goes to zero or unity.

\vspace{.5cm}
\subsection{Hybrid meson distribution amplitude}
\vspace{.5cm}

\noindent
In this subsection, we will consider in detail the properties of the hybrid meson
distribution amplitude (see also \cite{APSTW1}).
The Fourier transform of the hybrid meson --to--vacuum matrix element of
the bilocal vector quark operator may be written as
\begin{eqnarray}
&&\langle H(p,\lambda)| \bar \psi(-z/2)\gamma_\mu [-z/2;z/2]
\psi(z/2) |0 \rangle = \\
&&i f_H M_H \biggl[
\biggl( e_\mu^{(\lambda)}-p_{\mu}\frac{e^{(\lambda)}\cdot z}{p\cdot z} \biggr)
\int\limits_0^1 dy e^{i(\bar y - y)p\cdot z/2} \phi^{H}_T(y)+
 p_\mu \frac{e^{(\lambda)}\cdot z}{p\cdot z}
\int\limits_0^1 dy e^{i(\bar y - y)p\cdot z/2}\phi^{H}_{L}(y)\biggr],\nonumber
\label{hme}
\end{eqnarray}
where
$e^{(\lambda)}$ with $\lambda=(0,\,+1,\,-1)$
describes the polarization states of the hybrid meson. 
It is convenient to define the four-vector $e^{(0)}_L$
corresponding in the ultrarelativistic limit to the longitudinal polarization  as
\begin{eqnarray}
\label{polv}
e^{(0)}_{L\,\mu} = \frac{e^{(0)}\cdot z}{p\cdot z} p_\mu .
\end{eqnarray}

\noindent
For the longitudinal polarization case,
only the term with $\phi^{H}_{L}$ contributes, so that
\begin{eqnarray}
\langle H_L(p,0)| \bar \psi(-z/2)\gamma_\mu [-z/2;z/2]
\psi(z/2) | 0 \rangle =
i f_H M_H e^{(0)}_{L\,\mu}
\int\limits_0^1 dy e^{i(\bar y - y)p\cdot z/2} \phi^{H}_L(y)
\label{hmeW}
\end{eqnarray}
where $\bar y=1-y$ and $H$ denotes the isovector triplet of hybrid mesons;
$f_H$ denotes a dimensionful coupling constant of the
hybrid meson, so that $\phi^H$ is dimensionless. We will discuss its normalization later.

\noindent
In (\ref{hme}) and (\ref{hmeW}),
we insert the path-ordered gluonic exponential along the straight line connecting
the initial and final points $[z_1;z_2]$ which provides the gauge invariance
for bilocal operator and
equals unity in a light-like (axial) gauge. For  simplicity of notation we shall omit the index
$L$ from the  hybrid meson distribution amplitude.

\noindent
Although exotic quantum numbers like $J^{PC} = 1^{-+}$ are forbidden in the quark model,
it does not prevent the
leading twist correlation function from being non zero. The basis of the argument is that the
non-locality of the quark correlator opens the possibility of getting such a hybrid
state, because of dynamical gluonic degrees of freedom arising from the Wilson line.
This may be seen easily through a Taylor expansion of the non-local correlator
\begin{eqnarray}
\label{locdec}
&&\langle H(p,0)| \bar\psi(-z/2) \gamma_{\mu}[-z/2;z/2] \psi(z/2)| 0\rangle=
\\
&&\sum_{n\, odd}\frac{1}{n!}z_{\mu_1}..z_{\mu_n} \langle H(p,0)|
\bar\psi(0) \gamma_{\mu}
\stackrel{\leftrightarrow}{D}_{\mu_1}..\stackrel{\leftrightarrow}{D}_{\mu_n}
\psi(0)| 0\rangle ,
\nonumber
\end{eqnarray}
where $D_{\mu}$ is the usual covariant derivative and
$\stackrel{\leftrightarrow} {D_{\mu}}=\frac{1}{2}(
\stackrel{\rightarrow}{D_{\mu}}-
\stackrel{\leftarrow}{D_{\mu}}$).
The term corresponding to $z=0$ refers to the standard quark
model contribution, which is zero for the exotic hybrid quantum numbers. The first non
zero contribution arises from the first derivative contribution in this
expansion, and one can check that, more generally, only odd terms
contribute in this expansion. It is clear that due to gauge invariance,
such occurrence of operators $\stackrel{\leftrightarrow} {D_{\mu}}$
 naturally provides gluonic degrees of freedom, which enables
the production of hybrid state at twist 2 level. One can check explicitly
that the corresponding quantum numbers are indeed the one of the hybrid
state (see detailed discussion in \cite{APSTW1}).
Using charge conjugation invariance of $H^0,$ one can show that the
corresponding distribution amplitude is antisymmetric, namely
\begin{eqnarray}
\phi^{H}(y)=-\phi^{H}(1-y)\,.
\label{asy}
\end{eqnarray}
The result of this analysis is that the hybrid light-cone distribution amplitude is a
leading twist quantity
which should have a vanishing first moment  because of  
the antisymmetry property of the distribution
amplitude.
This distribution amplitude obeys usual non-singlet evolution equations \cite{ERBL}
and has an asymptotic limit \cite{Chase}
\begin{equation}
\Phi^H = 30  y (1-y) (1-2y)
\end{equation}
The  normalization factor (coupling constant) $f_{H}$ is defined through
the matrix element of
the energy-momentum tensor \cite{Koles}. It may be related, by making use
of the equations of motion, to the matrix element of quark-gluon operator
and  estimated with the help of the techniques of QCD sum-rules \cite{Balitsky}.
One of the solutions corresponds to a resonance with  mass
around $1.4\, {\rm GeV}$ and normalization factor\footnote{Our $f_{H}$
corresponds to $¥2 \sqrt{2} f_{R}$ in the notations of Ref. \cite{Balitsky}.}
\begin{equation}
 f_{H } \sim 50 MeV \,.
\end{equation}
If it turns out that only one resonance can be attributed to such an
hybrid state,
this QCD sum-rule analysis is sufficient to fix the value of $f_H.$ If
the scenario with two resonances is confirmed, this value of $f_H$
corresponds
to an effective coupling to the total contribution of these two
resonances. Despite the fact that QCD sum rules cannot distinguish at
the level of the coupling between two very close resonances and a single one, if one
would define $f_1$ (resp. $f_2$) the coupling to the first resonance
(resp. second), one should write
\begin{equation}
 f_{H }^2 = f_1^2 + f_2^2 \,.
\end{equation}
Thus, when selecting experimentally each resonance by their decay modes,
we know for sure that one of the coupling should be larger than
$f_H/\sqrt 2.$ Thus, the fact that the exotic hybrid quantum numbers could be attributed to two
very close resonances does not spoil the conclusion about the expected
order of magnitude of the hybrid distribution amplitude.
From now on, we will consider the case where only one of the $\pi_{1}¥$ candidates is
an exotic hybrid meson.

\noindent
The coupling constant $f_H$ is the subject of evolution given by the
formula, see e.g. \cite{DGP}
\begin{equation}
\label{evolution}
f_H(Q^2)= f_H \left(\frac{\alpha_S(Q^2)}{\alpha_S(M_H^2)}
\right)^{K_0}\;\;\;\;K_0=\frac{2\,\gamma_{QQ}(0)}{\beta_0}\;,
\end{equation}
where the anomalous dimension $\gamma_{QQ}(0)=16/9$ and
$\beta_0=11-2n_f/3$.
The exponent $K_0$ is thus
a small positive number which drives slowly to zero the coupling
constant $f_H(Q^2)$. Since experiments are likely to be feasible
at moderate values of $Q^2$, we neglect this evolution.

\vspace{.5cm}
\section{Cross-sections for hybrid meson electroproduction}
\vspace{.5cm}

\noindent
In this section, we focus on the computation and analysis of
the differential cross section for  longitudinally polarized hybrid meson 
electroproduction. 
The experimentally accessible differential cross
section corresponds to the process (\ref{pr}) for which the reaction (\ref{pr2})
is the subprocess. We assume that the virtual photon has
a longitudinal polarization, so that the factorization theorem is valid
owing to the absence of infrared divergences. We restrict ourselves to the
leading twist contributions.

\noindent
When we estimate hybrid meson electroproduction cross section, we systematically
compare it with the similar contribution ({\em i.e.} without gluon GPDs) to the 
cross section
for longitudinally polarized $\rho$ meson electroproduction.
The unpolarized cross section corresponding to the reaction  (\ref{pr2})
is defined by
\footnote{The flux factor is chosen as in \cite{Vand99}.}
\begin{eqnarray}
\label{xsecL}
\frac{d\sigma_L}{d\hat t}=\frac{1}{16\pi(\hat s-m_N^2)\lambda(\hat s,-Q^2,m_N^2)}
\frac{1}{2}\sum_{pol.} |{\cal A}_{(q)}|^2,
\end{eqnarray}
where the amplitude ${\cal A}_{(q)}$ is determined by (\ref{amp02});
$\hat s$, $\hat t$  are the usual Mandelstam variables
 and $m_N$ is the nucleon mass.
The function $\lambda$ is standardly defined by
\begin{eqnarray}
\label{lambda}
\lambda^2(x,y,z)=x^2+y^2+z^2-2xy-2xz-2yz.
\end{eqnarray}

\noindent
To calculate the cross section (\ref{xsecL}), we need to model the corresponding GPD's.
We apply the Radyushkin model \cite{Rad} where the function $H$, see (\ref{amp02}),
is expressed with the help of double distributions $F^q(x,y;t)$.
We have
\begin{eqnarray}
\label{GPD}
H^q(x,\xi,t)&=&\frac{\theta(\xi+x)}{1+\xi}
\int\limits_{0}^{{\rm min}\{ \frac{x+\xi}{2\xi},
\frac{1-x}{1-\xi} \}} dy\,
F^q\biggl(\frac{x+\xi-2\xi y}{1+\xi}, y, t\biggr) -
\nonumber\\
&&\frac{\theta(\xi-x)}{1+\xi}
\int\limits_{0}^{{\rm min}\{ \frac{\xi-x}{2\xi},
\frac{1+x}{1-\xi} \}} dy\,
F^{\bar q}\biggl(\frac{\xi-x-2\xi y}{1+\xi}, y, t\biggr).
\end{eqnarray}
where  standard notations are used.
For the double distribution $F^q(X,Y;t)$, we assume the ansatz suggested by Radyushkin 
\cite{Rad}:
\begin{eqnarray}
\label{DD}
F^q(X,Y;t)=\frac{F_1^q(t)}{F_1^q(0)}\, q(X)\, 6\frac{Y(1-X-Y)}{(1-X)^3},
\end{eqnarray}
and a similar expression for the anti-quark contribution.
As shown in \cite{PW},
this  definition of the double distribution is not completely compatible with
the structure of the corresponding matrix elements; introducing D-terms
restores the self-consistency of this representation. Taking into account these D-terms,
the GPD's (\ref{GPD}) are modified into :
\begin{eqnarray}
\label{eqnarray}
H^q_D(x,\xi,t)=H^q(x,\xi,t)+\theta(\xi-|x|)\frac{D(x/\xi, t)}{N_f},
\end{eqnarray}
where $D(x/\xi,0)$ is given as in \cite{L-D}.
In the present paper, we concentrate on the region where
the values of the skewedness parameter $\xi$ are rather small. Hence, it is legitimate to
neglect these D-terms in the amplitude of  $\rho$ meson production.
Meanwhile their contributions to the hybrid meson production  always
vanish owing to the anti-symmetric properties of these D-terms.

\begin{figure}
$$\rotatebox{270}{\includegraphics[width=8cm]{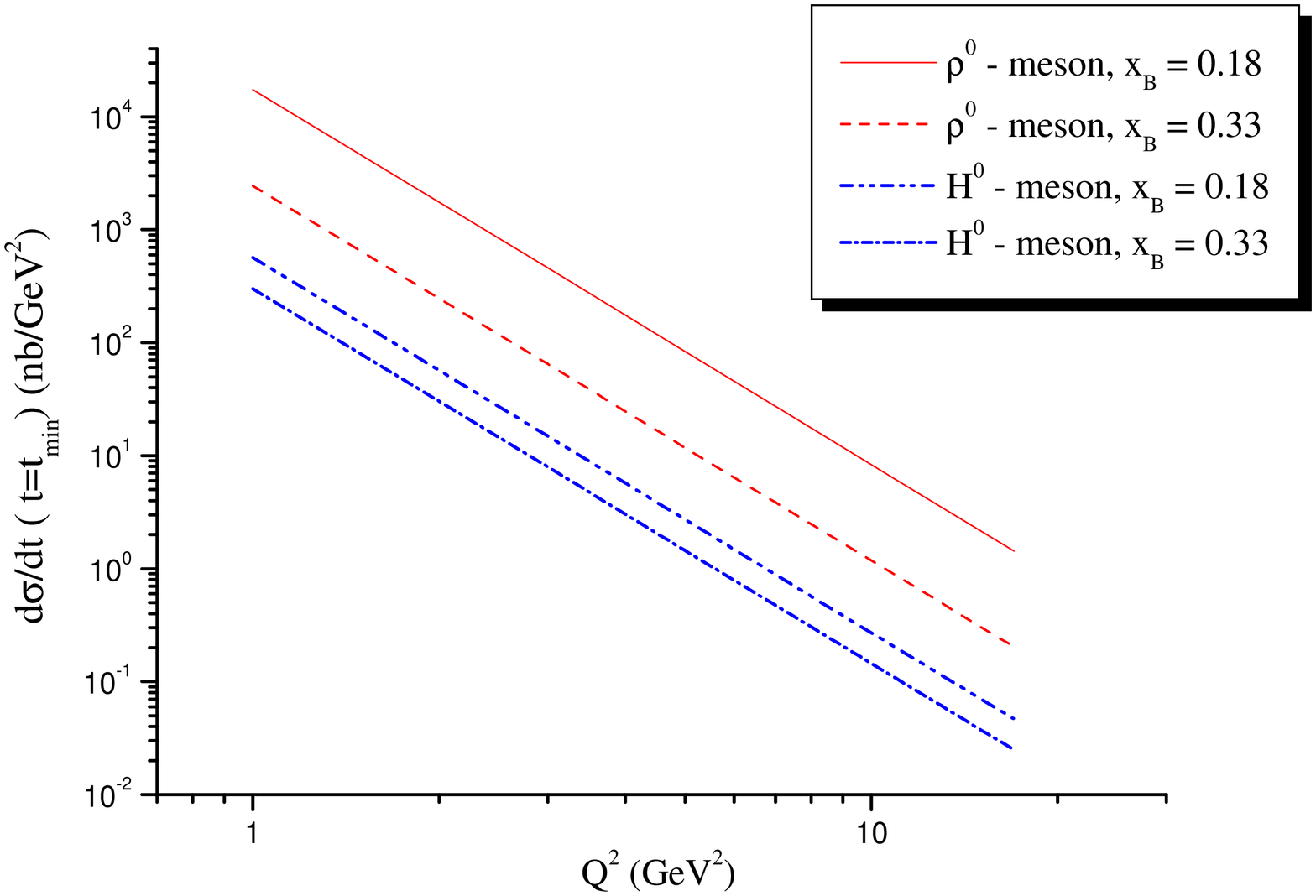}}$$
\caption{Differential cross section for $\rho$ and hybrid meson production
 with the naive choice of the renormalization scale and different $x_B$.}
\label{conaiv}
\end{figure}

\begin{figure}
$$\rotatebox{270}{\includegraphics[width=8cm]{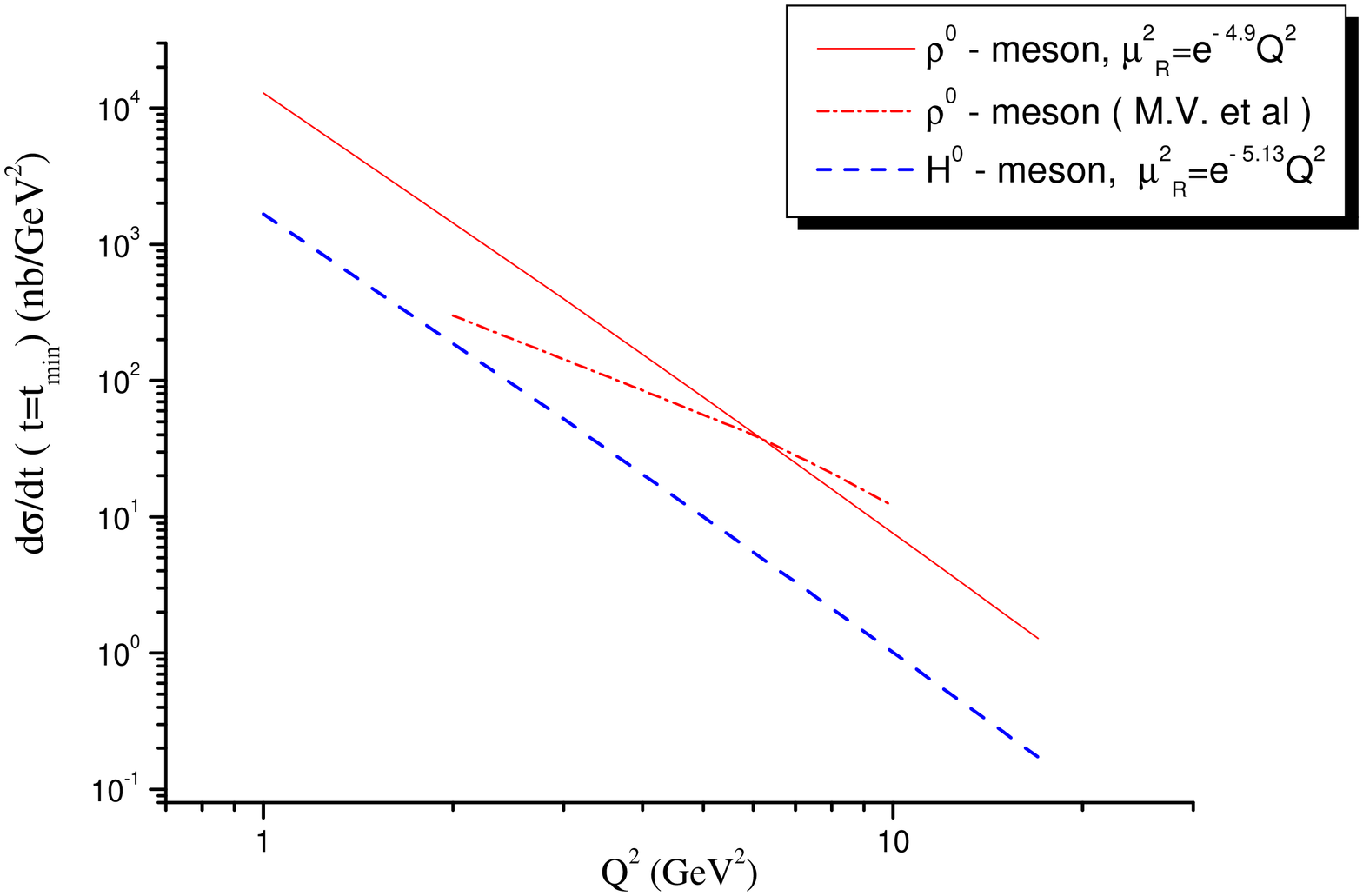}}$$
\caption{Differential cross-section for exotic hybrid meson
electroproduction (dashed line) with
$\mu_R^2 =e^{-5.13} Q^2$ compared with
the quark contribution to $\rho^0$
electroproduction (solid line) with $\mu_R^2 =e^{-4.9} Q^2$, as a function
of $Q^2,$ for $x_B\approx 0.33$. The dash-dotted line is the result of Vanderhaegen et al 
\cite{Vand99} for $\rho$ electroproduction.}
\label{xsec2}
\end{figure}

\noindent
In (\ref{DD}), functions $q(x)$ and $\bar q(x)$ are the ordinary quark and anti-quark
distributions in the nucleon for which we use the MRST98 parameterization
\cite{MRST98}. An important aspect of each model of GPD's is its dependence ot $t$ \cite{Burk}.
Here it is assumed to be factorizable through the
 functions $F_1^q(t)$ for each flavour, which are  equal to
\begin{eqnarray}
F_1^u=2F_1^p+F_1^n, \quad F_1^d=2F_1^n+F_1^p ,
\end{eqnarray}
where $F_1^p$ and $F_1^n$ are the proton and neutron electromagnetic form factors.
Note that we neglected the strange form factor because it is small.
In the same way, we can write the expression for the function $E$.
We neglect here its contribution because it is small and quite model-dependent.

\noindent
Finally, to get prediction for the cross sections we need to fix the renormalization 
scales. In order to estimate theoretical uncertainties of this procedure we 
fix the scale $\mu_R^2$ in two different ways: firstly, in the naive way, by assuming 
$\mu^2_R=Q^2$, and secondly, by applying the BLM prescription \cite{BLM}.

\noindent
The resulting differential cross sections for hybrid meson and $\rho$ meson (quark
contribution only) production  are shown on Fig. \ref{conaiv} for $x_B = 0.18$ and 
$0.33$, using the above mentioned naive scale fixing.

\noindent
The BLM procedure, which is discussed in details in \cite{APSTW3},
leads to the following values of the renormalization scales:
\begin{eqnarray}
\label{BLMsc1}
&&\mu^2_R=e^{-4.9}Q^2, \quad \mbox{for\,\, $\rho$ \,\,meson}, 
\nonumber \\
&&\mu^2_R=e^{-5.13}Q^2, \quad \mbox{for\,\, $H$\,\, meson.} 
\end{eqnarray}
for the case $\xi=0.2$ (or $x_B\approx 0.33$), and
\begin{eqnarray}
\label{BLMsc2}
&&\mu^2_R=e^{-4.68}Q^2, \quad \mbox{for\,\, $\rho$ \,\,meson}, 
\nonumber \\
&&\mu^2_R=e^{-5.0}Q^2, \quad \mbox{for\,\, $H$\,\, meson.} 
\end{eqnarray}
for the case $\xi=0.1$ (or $x_B\approx 0.18$).

\noindent
Note that, taking into account the D-terms, the $\rho$ meson BLM scale
is slightly diminished. For instance, in the case $x_B\approx 0.33$ we have
\begin{eqnarray}
\label{r02D}
\mu^2_R = \, e^{-5.4}\, Q^2 .
\end{eqnarray}

\noindent
These renormalization scales have rather small
magnitudes. This has a tendency to enlarge the cross sections but may endanger the validity
of the perturbative approach. However, it is possible that
the coupling constant $\alpha_S$ stays below unity and 
the perturbative theory does not suffer from the IR divergencies.
We will use the Shirkov and Solovtsov's ansatz \cite{Shirkov}
where the analytic running coupling constant takes the form:
\begin{eqnarray}
\label{asan}
\alpha_{S}^{an}(\mu_R^2) = \frac{4\pi}{\beta_0}
\biggl[
\frac{1}{{\rm ln} \mu_R^2/\Lambda_{QCD}^2}+\frac{\Lambda_{QCD}^2}{\Lambda_{QCD}^2-\mu^2_R}
\biggr].
\end{eqnarray}
Here $\Lambda_{QCD}$ is the standard scale parameter in QCD.
The second term in (\ref{asan}) assures the absence of a ghost pole at $\mu_R^2=\Lambda_{QCD}^2$
and has a nonperturbative source. Detailed discussion on this point may be found in
\cite{Bakulev} and references therein.

\noindent
Recently, in \cite{Vand99} the role of power 
corrections due to the intrinsic transverse momentum
of partons (the kinematical higher twist) has been investigated.
In that approach the inclusion of the intrinsic transverse momentum
dependence results in a rather strong effect on the differential cross-section  
before the scaling regime is achieved. In \cite{Vand99}, the renormalization 
scale $\mu^2_R$ is defined by
the gluon virtuality so that the scale is a function of parton fractions
flowing into the corresponding gluon propagator. 

\noindent
On Fig. \ref{xsec2}, we present our results for the differential cross section of the hybrid 
meson electroproduction compared to the $\rho$ meson electroproduction, using the BLM scales. 
We can see that the hybrid cross section is rather sizeable in comparison with 
the corresponding $\rho$ meson cross section.
We also show the results obtained in \cite{Vand99} for the $\rho$ meson electroproduction. 
We see that in the region $Q^2\sim 5 - 10\, {\rm GeV}^2$ the size of
the $\rho$ meson cross section obtained  with the inclusion of  
transverse momentum effects 
is very close to the analogous cross section computed with the BLM scale and without
the intrinsic transverse momentum dependence. On the other hand, for higher values of $Q^2$
 the leading order amplitude 
computed with the BLM scale fixing  is falling faster that the 
corresponding amplitude derived in Ref. \cite{Vand99}, whereas for smaller 
values of $Q^2$ it is larger than that prediction.
We do not want to claim here that  kinematical higher twist contributions
have no effects at  low values of $Q^2$ but rather that a rather strong
effect on the $Q^2$ dependence
of the cross sections may be dictated by another mechanism which is much more
controllable since it depends on the estimate of higher order perturbative contributions.

\noindent
All this shows that the scale fixing ambiguities lead to a non negligible theoretical 
uncertainty on the absolute value of cross sections. It is important 
however to understand that most of this uncertainty
does not apply to ratios of cross sections, and in particular to the most interesting 
ratio 
$d\sigma^H \,: \,d\sigma^{\rho}$, which measures the expected cross section for hybrid
production 
with respect to the well measured and large cross section for $\rho$ meson production. 
Indeed, as shown on Table $1$, this ratio is very insensitive to the scale fixing procedure. 
Moreover it is not small when $x_B$ is 
large enough and almost $Q^2$ independent. The decreasing value of the 
ratio when $x_B$ diminishes comes from the relative sign of the two terms 
contributing in (\ref{amp02}), {\it i.e.} when $\xi \to 0$ the structure
$\cal{H}^-$ goes to zero too.

\noindent
In conclusion of this section, we would like to stress that
the present work has demonstrated the feasibility of hybrid meson production experiments
in electroproduction at moderate energies. An obvious remaining question is how much of
the cross section is observable in a dedicated experiment. If the experiment is able to detect 
the final state electron and baryon and to measure their momenta with good accuracy, a missing
mass analysis may allow to identify and study all decay channels of the hybrid. In the next 
sections, we discuss the cases where the 
hybrid meson is detected through a particular decay channel.

\begin{center}
\begin{tabular}{|r||l|l|l|l||l|l|l|l|} 
\hline
$x_B$&\multicolumn{4}{|c||}{$0.33$} &\multicolumn{4}{|c|}{$0.18$} \\
\hline\hline
 & & & & & & & & \\
$Q^2 \,({\rm GeV}^2)$& 3.0&7.0&11.0&17.0&3.0&7.0&11.0&17.0\\ 
\hline
 & & & & & & & & \\
$\mu^2_R =Q^2$&0.123&0.123&0.123&0.123&0.0325&0.0326&0.0326&0.0326\\
\hline
 & & & & & & & & \\
$ \mu^2_R =\mu_{BLM}^2$&0.131&0.133&0.133&0.134&0.0356&0.0362&0.0365&0.0367\\
\hline
\end{tabular}

\vspace{.5cm}
Table 1: Ratio $d\sigma^H : d\sigma^{\rho}$ for both the naive and BLM scales and
for the different values of $x_B$. 
\end{center}

\vspace{.5cm}
\section{Study of hybrid mesons via the electroproduction of $\pi\eta$ pairs}
\vspace{.5cm}

\begin{figure}
$$\includegraphics[width=12cm]{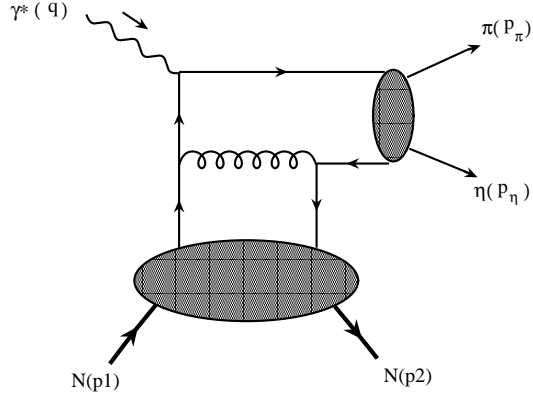}$$
\caption{Typical diagram describing the electroproduction of
$\pi\eta$ pair. The higher and lower blobs represent the GDA's and GPD's, respectively.}
\label{diagrampieta}
\end{figure}

\noindent
In the case where there is no recoil detector which allows to identify the
hybrid production events through a missing mass reconstruction, one will have to base
an identification process through the possible decay products of the hybrid meson $H^0$. Since the
particle $\pi_{1} (1400)$  has a dominant $\pi\eta $ decay mode, we now
proceed to the description
\footnote{A very similar analysis may be carried for the $\pi \eta'$ decay mode of
the candidate $\pi_{1} (1600).$} of the electroproduction  process
\begin{eqnarray}
\label{eN}
e(k_1)+N(p_1)\to e(k_2)+\pi^0(p_\pi)+\eta(p_\eta) + N(p_2)
\end{eqnarray}
or
\begin{eqnarray}
\label{exreac}
\gamma^*(q)+N(p_1)\to \pi^0(p_\pi)+\eta(p_\eta) + N(p_2).
\end{eqnarray}
To perform a leading order computation of such process (see Fig. \ref{diagrampieta}).
we need to introduce  the concept of generalized distribution
amplitude (GDA) \cite{DGPT} for $\pi\eta$.

\vspace{.5cm}
\subsection{$\pi\eta$ generalized distribution amplitude}
\vspace{.5cm}

\noindent
In this subsection, we briefly introduce and  discuss the generalized distribution amplitude
related to the $\pi\eta$--to--vacuum matrix element. On the basis of  Lorentz
invariance, the $\pi^0\eta$ GDA may be defined
\footnote{A straightforward generalization enables to
write similar equations for charged states} as :
\begin{eqnarray}
\label{hme2}
&&\langle \pi^0(p_\pi)\eta(p_\eta) |
\bar\psi_{f_2}(-z/2)\gamma^{\mu}[-z/2;z/2] \tau^3_{f_{2}f_{1}}
\psi_{f_1}(-z)|0\rangle=
\nonumber\\
&&p^{\mu}_{\pi\eta}\int\limits_{0}^{1}dy e^{i(\bar y-y)p_{\pi\eta}\cdot z/2}
\Phi^{(\pi\eta)}(y,\zeta, m_{\pi\eta}^2),
\end{eqnarray}
where  the total momentum of $\pi\eta$ pair is $p_{\pi\eta}=p_{\pi}+p_{\eta}$
while $m^2_{\pi\eta}=p^2_{\pi\eta}$.
We omit the $Q^2$ dependence of the $\pi^0\eta$ GDA's which is the same as the 
one discussed above for the hybrid meson distribution amplitude.
Note that the $\pi\eta$ distribution amplitude $\Phi^{(\pi\eta)}$ describes non resonant
as well as resonant contributions. It does not possess any symmetry properties
concerning the $\zeta$-parameter.

\noindent
Let us now discuss the $\zeta$ parameter.
When the two mesons have equal masses, the parameter $\zeta$ is usually defined as
$\zeta=p_{\pi}^+ / p^+$.
In the case of two different particles it is more convenient
to define the parameter $\tilde\zeta$ in the following way:
\begin{eqnarray}
\label{zeta}
&&\tilde\zeta=\frac{p_\pi^+}{(p_\pi+p_\eta)^+}-
\frac{m^2_\pi-m^2_\eta}{2m^2_{\pi\eta}},
\nonumber\\
&&1-\tilde\zeta=\frac{p_\eta^+}{(p_\pi+p_\eta)^+}+
\frac{m^2_\pi-m^2_\eta}{2m^2_{\pi\eta}}.
\end{eqnarray}
Then,  we  get the ordinary relation between $\tilde\zeta$  and the angle $\theta_{cm}¥$,
defined as the polar angle of the $\pi$ meson in the center of mass frame of the meson pair:
\begin{eqnarray}
\label{2zeta}
2\tilde\zeta-1=\beta\cos\theta_{cm}.
\end{eqnarray}
In (\ref{2zeta}), the standard $\beta$-function is given by
\begin{eqnarray}
\beta=\frac{2|{\bf p}|}{m_{\pi\eta}},
\end{eqnarray}
where ${\bf |p|}$ denote the modulus of three-dimension momentum of $\pi$ and $\eta$
mesons in the center--of--mass system.

\noindent
In the reaction under  study, the $\pi\eta$ state may have
total momentum, parity and charge-conjugation in the following
sequence
$$J^{PC}=0^{++},\, 1^{-+},\, 2^{++}, \, ...$$
that corresponds to the following values of the $\pi\eta$ orbital angular momentum $L$:
$$L=0,\, 1, \, 2,\, ...$$
respectively. We can see that a resonance with a $\pi\eta$ decay mode for odd
orbital angular momentum $L$ should be considered as an exotic meson.

\noindent
The mass region around $1400$ ${\rm MeV}$ is dominated by the strong $a_2(1329)\,(2^{++})$
resonance \cite{Adams}. It is therefore natural to look for the
interference of the amplitudes of hybrid and $a_{2}$ production, which is linear, rather
than quadratic in the hybrid electroproduction amplitude.
Such interference arises from the usual representation of the $\pi\eta$ generalized
distribution amplitude
in the form suggested by its asymptotic expression :
\begin{eqnarray}
\label{asPhi}
\Phi^{(\pi\eta),\,a}(y,\tilde\zeta, m_{\pi\eta}^2)=10 y (1-y)
C^{(3/2)}_1(2y-1)
\sum_{l=0}^{2} B_{1l}(m_{\pi\eta}^2) P_l(\cos\theta).
\end{eqnarray}
Keeping only $L=1$ and $L=2$ terms,
we model the $\pi\eta$ distribution amplitude in the following form:
\begin{eqnarray}
\label{approx}
\Phi^{(\pi\eta)}(y,\zeta, m_{\pi\eta}^2)=30 y (1-y)(2y-1)
\biggl[
\,B_{11}(m_{\pi\eta}^2) P_1(\cos\theta) +
B_{12}(m_{\pi\eta}^2)
P_2(\cos\theta)
\biggr],
\end{eqnarray}
with the coefficient functions $B_{11}(m_{\pi\eta}^2)$ and
$B_{12}(m_{\pi\eta}^2)$   related to corresponding
Breit-Wigner amplitudes when
$m^2_{\pi\eta}$ is in the vicinity of $ M^2_{a_2},\,M^2_H$. We have
(see the technical details how to calculate these coefficient functions in
Appendix A):
\begin{eqnarray}
\label{B11-2}
 B_{11}(m^2_{\pi\eta})\biggl|_{m^2_{\pi\eta}\approx M^2_H}=
\frac{5}{3}\,
\frac{g_{H\pi\eta}f_H M_H \beta}{M^2_H-m^2_{\pi\eta}-i\Gamma_H M_H}
\end{eqnarray}
and
\begin{eqnarray}
B_{12}(m^2_{\pi\eta})\biggr|_{m^2_{\pi\eta}\approx M^2_{a_2}}=
\frac{10}{9} \frac{i g_{a_2\pi\eta} f_{a_2} M^2_{a_2} \beta^2}
{M^2_{a_2}-m^2_{\pi\eta}-i\Gamma_{a_2} M_{a_2}}.
\end{eqnarray}

\noindent
In the $a_{2}$ case, we use the results and conventions of \cite{BraunKivel};
note that  the coupling constant $g_{a_{2}\pi\eta}$ has mass dimension
equal to $-1$.

\noindent
The coupling constants $g_{H\pi\eta}$ and $g_{a_2\pi\eta}$
 may be estimated through the
approximate measurements of the partial widths of the $a_2$ and hybrid meson  in the
$\pi\eta $ decay channel, we have:
\begin{eqnarray}
\label{GHa}
&&\Gamma(H\to\pi\eta)=\frac{1}{16\pi}g^2_{H\pi\eta}
\frac{\lambda^3(M^2_H, m^2_\pi, m^2_\eta)}{M^5_H},
\nonumber\\
&&\Gamma(a_{2}\to\pi \eta)=\frac{1}{24\pi}g^2_{a_2\pi\eta}
\frac{\lambda^5(M^2_{a_2}, m^2_\pi, m^2_\eta)}{M^7_{a_2}}.
\end{eqnarray}
Neglecting the masses of $\pi$ and $\eta$ mesons compared to the
$a_2$ and hybrid meson masses,
we get
\begin{eqnarray}
\label{gHpieta}
g^2_{H\pi\eta}\approx
\frac{16\pi}{M_H}\Gamma(H\to\pi\eta), \quad
g^2_{a_2\pi\eta}\approx
\frac{24\pi}{M_{a_2}^3}\Gamma(a_2\to\pi\eta).
\end{eqnarray}

\vspace{.5cm}
\subsection{Differential cross section for $\pi\eta$ electroproduction}
\vspace{.5cm}

\begin{figure}
$$\includegraphics[width=12cm]{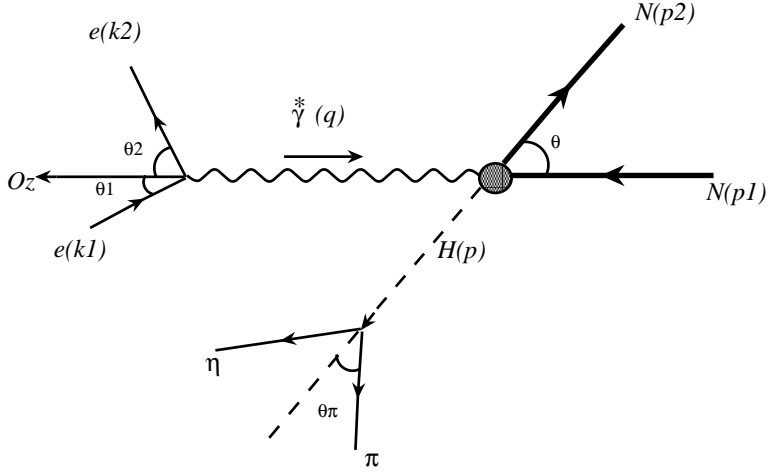}$$
\caption{Typical process describing the electroproduction of a
$\pi\eta$ pair. }
\label{diagpr}
\end{figure}

\noindent
Let us first  fix the kinematics. For reaction (\ref{eN}) we choose:
\begin{eqnarray}
\label{genkin}
&&k_1=(\varepsilon_1,\, \varepsilon_1\sin\theta_1,\,0,\,\varepsilon_1\cos\theta_1)
\nonumber\\
&& p_1=(E_1,\,0,\,0,\,p^3_1), \quad q=k_1-k_2=(q_0,\,0,\,0,\,-p^3_1),
\nonumber\\
&& p_2=(E_2,\,|{\bf p}_2|\cos\phi\sin\theta,\,
|{\bf p}_2|\sin\phi\sin\theta,\,|{\bf p}_2|\cos\theta).
\end{eqnarray}
The following Mandelstam and dimensionless variables can be defined as
(here, the "hatted" symbols refer to the subprocess (\ref{exreac}))
\begin{eqnarray}
\label{Mvs}
\hat s=(p_{\pi\eta}+p_2)^2=(q+p_1)^2, \quad \hat t=(p_2-p_1)^2, \quad S=(k_1+p_1)^2,
\end{eqnarray}
and
\begin{eqnarray}
x_B=\frac{Q^2}{2p_1\cdot q}, \quad y_l=\frac{p_1\cdot q}{p_1\cdot k_1}.
\end{eqnarray}

\noindent
In (\ref{genkin}), the energies and momenta can be expressed as
\footnote{Here,  $\varepsilon_2$ is the energy of the scattered lepton.}
\begin{eqnarray}
&&\varepsilon_1=\frac{S-m_N^2-Q^2}{2\sqrt{\hat s}}, \quad
\varepsilon_2=\frac{S-\hat s}{2\sqrt{\hat s}},
\nonumber\\
&&E_1=\frac{\hat s +m^2_N+Q^2}{2\sqrt{\hat s}}, \quad
E_2=\frac{\hat s -m^2_{\pi\eta}+m^2_N}{2\sqrt{\hat s}}, \quad
q_0=\frac{\hat s -m^2_N-Q^2}{2\sqrt{\hat s}},
\nonumber\\
&&|{\bf q}|=|{\bf p}_1|=\frac{\lambda(\hat s,m^2_N,-Q^2)}{2\sqrt{\hat s}}, \quad
|{\bf p}_{\pi\eta}|=|{\bf p}_2|=\frac{\lambda(\hat s,m^2_{\pi\eta},m^2_N)}{2\sqrt{\hat s}},
\end{eqnarray}
where the kinematical function $\lambda$ in defined in (\ref{lambda}).

\noindent
The corresponding angles take the  forms
\footnote{ $\theta_2$ defines the polar angle of the final lepton.}
\begin{eqnarray}
\label{ang}
&&\cos\theta_2=\frac{2Q^2\hat s}{(S-\hat s)\lambda(\hat s, m^2_N,-Q^2)}-
\frac{\hat s -m^2_N-Q^2}{\lambda(\hat s,m^2_N,-Q^2)},
\nonumber\\
&&\cos\theta=
\frac{2\hat s(\hat t-2m^2_N)+(\hat s+m^2_N+Q^2)(\hat s-m^2_{\pi\eta}+m^2_N)}
{\lambda(\hat s, m^2_N,-Q^2)\lambda(\hat s, m^2_{\pi\eta},m^2_N)}.
\end{eqnarray}

\noindent
It is useful to note the following relations between the invariants:
\begin{eqnarray}
&&x_B=\frac{Q^2}{\hat s +Q^2-m^2_N},  \quad y_l=\frac{Q^2}{x_B(S-m^2_N)}=
\frac{\hat s+Q^2-m^2_N}{S-m^2_N},
\nonumber\\
&&Q^2=x_B y_l(S-m^2_N),\quad \hat s=\frac{1-x_B}{x_B}Q^2+m^2_N.
\end{eqnarray}

\noindent
One may also work within the
center-of-mass system of the meson pair, where we have after the corresponding boost,
\begin{eqnarray}
\label{pietacm}
p_\pi=(E_\pi,\,|{\bf p}|\sin\theta_{cm},\,0,\,|{\bf p}|\cos\theta_{cm}), \quad
p_\eta=(E_\eta,\,-|{\bf p}|\sin\theta_{cm},\,0,\,-|{\bf p}|\cos\theta_{cm}),
\end{eqnarray}
where the energies and momenta of the mesons take the forms
\begin{eqnarray}
E_\pi=\frac{m^2_{\pi\eta}-m^2_\eta+m^2_\pi}{2m_{\pi\eta}}, \quad
E_\eta=\frac{m^2_{\pi\eta}-m^2_\pi+m^2_\eta}{2m_{\pi\eta}},  \quad
|{\bf p}|=\frac{\lambda(m^2_{\pi\eta},m^2_\eta,m^2_\pi)}{2m_{\pi\eta}}.
\end{eqnarray}

\noindent
We now come to the expression for the differential cross section of
reaction (\ref{eN}).
The amplitude of this reaction is given by
\begin{eqnarray}
\label{genam}
T^{\pi^0\eta}=\bar u(k_2,s_2)\gamma\cdot\varepsilon_L u(k_1,s_1)
\frac{1}{q^2}{\cal A}_{(q)}^{\pi^0\eta },
\end{eqnarray}
leading, keeping the leading order in $Q^2$, to
\begin{eqnarray}
\label{sqT}
|T^{\pi^0\eta}|^2=
\frac{4e^2(1-y_l)}{Q^2 y_l^2}
|{\cal A}_{(q)}^{\pi^0\eta }|^2\,.
\end{eqnarray}
The amplitude of subprocess (\ref{exreac})
reads
\begin{eqnarray}
\label{qdsim3}
&&{\cal A}_{(q)}^{\pi^0\eta }=
\frac{e\pi\alpha_s C_F}{N_c\,Q}
\biggl[ e_u {\cal H}_{uu} -e_d {\cal H}_{dd}\biggr]
\nonumber\\
&&\biggl[ B_{11}(m_{\pi\eta}^2) P_1(\cos\theta_{cm}) +
B_{12}(m_{\pi\eta}^2)P_2(\cos\theta_{cm})
\biggr].
\end{eqnarray}

\noindent
Finally, the differential cross section of process (\ref{eN})  takes
the form
\begin{eqnarray}
\label{xsec}
\frac{d\sigma^{\pi^0\eta}}{dQ^2\, dy_l\,d\hat t\,
dm_{\pi\eta}\, d(\cos\theta_{cm})}
=\frac{1}{4(4\pi)^5}\,
\frac{m_{\pi\eta}\beta}{y_l\lambda^2(\hat s,-Q^2,m^2_N)}
\,|T^{\pi^0\eta}|^2.
\end{eqnarray}

\vspace{.5cm}
\section{Calculation of the angular asymmetry}
\vspace{.5cm}

\noindent
\begin{figure}
$$\rotatebox{270}{\includegraphics[width=8cm]{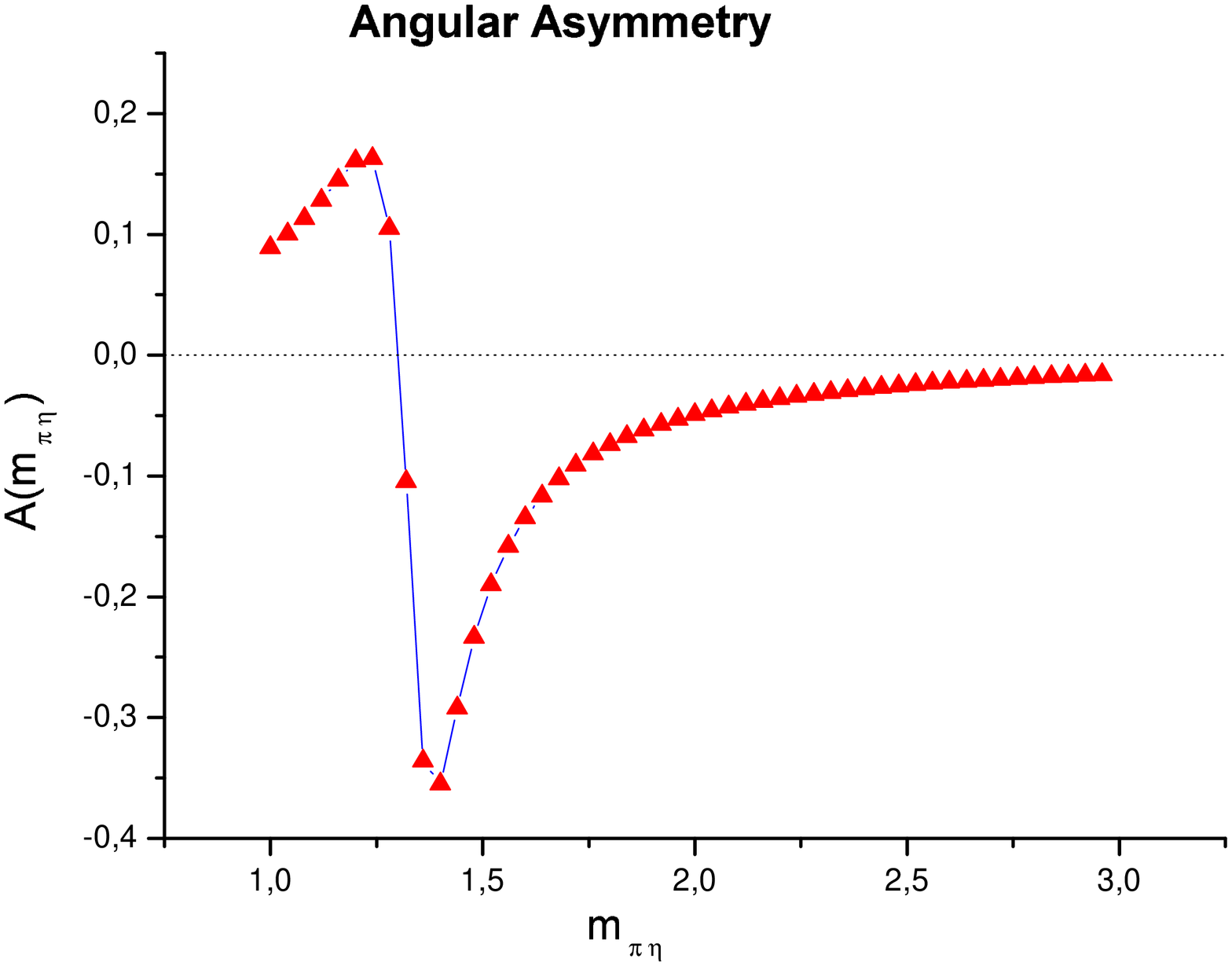}}$$
\caption{The angular asymmetry as a function of $m_{\pi\eta}$.}
\label{angular}
\end{figure}

\noindent
Asymmetries are often a good way to get a measurable signal for a small
amplitude, by taking profit of its interference with a larger one. In our case,
since the hybrid production amplitude may be rather small with respect to
a continuous background, we propose to use the supposedly
large amplitude for $a_{2}$ electroproduction as a magnifying lens to unravel
the presence of the exotic hybrid meson. Since these two amplitudes describe
different orbital angular momentum of the $\pi$ and $\eta$ mesons, the asymmetry
which is sensitive to their interference is an angular asymmetry  defined by
\begin{eqnarray}
\label{anas}
 A(Q^2, y_l,\hat t, m_{\pi\eta})=
 \frac{\int \cos\theta_{cm} \,
d\sigma^{\pi^0\eta}(Q^2, y_l,\hat t, m_{\pi\eta}, \cos\theta_{cm} )}{
\int d\sigma^{\pi^0\eta}
(Q^2, y_l,\hat t, m_{\pi\eta}, \cos\theta_{cm} )}
\end{eqnarray}
as a weighted integral over polar angle $\theta_{cm}$ of the relative momentum of
$\pi$ and $\eta$ mesons.
The angle  $\theta_{cm}$ is related to the  parameter $\tilde \zeta$ by  formula (\ref{2zeta}).
Due to the fact that the $\cos\theta_{cm}$-independent factors in both
the numerator and denominator of (\ref{anas}) are completely factorized and,
on the other hand, these factors are the same, we are able to rewrite the asymmetry (\ref{anas})
as
\begin{eqnarray}
\label{anas2}
 A(m_{\pi\eta})=
\frac{\int d(\cos\theta_{cm}) \, \cos\theta_{cm}
\biggl| B_{11}(m_{\pi\eta}^2) P_1(\cos\theta_{cm}) +
B_{12}(m_{\pi\eta}^2)P_2(\cos\theta_{cm})
\biggr|^2}{
\int d(\cos\theta_{cm}) \,
\biggl| B_{11}(m_{\pi\eta}^2) P_1(\cos\theta_{cm}) +
B_{12}(m_{\pi\eta}^2)P_2(\cos\theta_{cm})
\biggr|^2},
\end{eqnarray}
and, calculating the $\cos\theta_{cm}$-integral analytically,
to obtain
\begin{equation}
\label{anglas}
A( m_{\pi\eta} )=\frac{N( m_{\pi\eta})}
{D ( m_{\pi\eta} )},
\end{equation}
with
\begin{equation}
N = \frac{8}{15} \Re
\biggl[ B_{11}(m_{\pi\eta}^2) B_{12}^*(m_{\pi\eta}^2) \biggr],
\quad
D = \frac{2}{3} \biggl| B_{11}(m_{\pi\eta}^2)\biggr|^2+
\frac{2}{5}
\biggl| B_{12}(m_{\pi\eta}^2)\biggr|^2 .
\end{equation}
While in the two-pion production case the interference between
the isoscalar and isovector channels can be investigated, we here restrict to
 the interference between $L=1$ and $L=2$ modes of $\pi^0\eta$.
As a result, the introduction of the so-called intensity density (see \cite{L-D}),
{\it i.e.} the integrated-over-invariants value,  useful for the two-pion modes,
 completely coincides with the value (\ref{anglas}) for our case.

\noindent
Our estimation of the asymmetry (\ref{anglas}) is shown on Fig. \ref{angular}.
Since the numerator of (\ref{anglas}),
{\it i.e.} the real part of the product of $B_{11}(m_{\pi\eta}^2)$ and
$B_{12}^*(m_{\pi\eta}^2)$, is proportional to the cosine of the phase difference
$\Delta\delta_{1,2}=\delta_{l=1}-\delta_{l=2}$ the zeroth
value of (\ref{anglas})  takes place at $\Delta\delta_{1,2}=\pi/2$.
This is achieved for $m_{\pi\eta}\approx 1.3 \, {\rm GeV}$.
Besides, one can see from Fig. \ref{angular} that the first positive extremum
is located at $m_{\pi\eta}$ around the mass of $a_2$ meson while the second negative extremum
corresponds to the hybrid meson mass.

\noindent
Note that this angular asymmetry  is completely similar to the
charge asymmetry which was studied in $\pi^+\pi^-$ electroproduction
at HERMES \cite{hermes}.

\vspace{0.5cm}
\section{Note on the $\pi\pi\pi$ channel}
\vspace{0.5cm}

\noindent
The hybrid candidate $\pi_{1}(1600)$ has been also seen through a $\pi\pi\pi$ decay
channel. In that case the deep exclusive electroproduction of three pions provides
the background which should be studied, including the possible interference effects of hybrid
meson and the background. The analysis that we have described in section 4 and 5 may be adapted
to the three body case by using the results of \cite{PT}. At the leading twist level
the generalized distribution amplitude  of the charge conjugation even state may be written in
complete analogy to the pion light-cone  distribution from the large distance matrix element
\begin{eqnarray}
\label{soft-S}
S_{\alpha\beta} = {P^+ \over2\pi} \int\! dx^-\,
  e^{-i y (P^+ x^-)} \,
\out\langle \pi \pi \pi |\,
\bar{\psi}_\alpha(x^- v')\psi_\beta(0) \,| 0 \rangle\in
\end{eqnarray}
as
\begin{equation}
S_{q, \alpha \beta}^{\phantom{\mu}}\,  \gamma^{+}_{\alpha \beta} \gamma_5
   =   \frac{i}{f_\pi}\Phi^+_q (z, \zeta_0, \zeta_+,\zeta_-;
W_{12}^2,  W_{13}^2,  W_{23}^2,) \, P^+ .
\end{equation}
The three light cone fractions are normalized by the condition
$\zeta_0+ \zeta_+ +\zeta_- =1$, making only two of them independent,
while the squared total energy of the three pions is
$W^2=W_{12}^2+W_{13}^2+W_{23}^2 -3m_\pi^2$, where $W_{ij}$ are the invariant masses
of the pairs of mesons. 

\noindent
Charge conjugation invariance provides a symmetry
relation:
\begin{equation}
\label{symmetry}
\Phi^+(y, \zeta_0, \zeta_+,\zeta_-) = \Phi^+(1-y, \zeta_0,
\zeta_-, \zeta_+).
\end{equation}
The asymptotic $z$-dependence is just
\begin{equation}
\label{symmetry2}
\Phi^+_q (y, \zeta_0, \zeta_+,\zeta_-) = \frac{1}{6(1+a)}
y (1-y) (\zeta_0+a \zeta_0^2),
\end{equation}
where $a$ is an unknown parameter and the normalization has been fixed with
the help of the fact that putting both charged pion momenta to zero,
one should get the GDA  equal to the pion distribution amplitude. The QCD evolution
 is the same as for the pion distribution amplitude, {\em i.e.} with a  vanishing anomalous
 dimension.

\noindent
Conversely, the generalized distribution amplitude  $\Phi^-(y, \zeta_0, \zeta_+,\zeta_-)$
 of the charge conjugation odd state obeys the equation :
 \begin{equation}
\Phi^-(y, \zeta_0, \zeta_+,\zeta_-) =- \Phi^-(1-y, \zeta_0,
\zeta_-, \zeta_+).
\end{equation}
and its asymptotic $z$-dependence is
\begin{equation}
\Phi^-_q (y, \zeta_0, \zeta_+,\zeta_-) \sim
y (1-y) (2y-1)P(\zeta_0, \zeta_+,\zeta_-),
\end{equation}
where $P$ is a polynomial of degree 3 with the symmetry property
$$P(\zeta_0, \zeta_+,\zeta_-) = P(\zeta_0, \zeta_-,\zeta_+).$$
Its QCD evolution is the same as for the hybrid distribution amplitude.
The interference with $\pi_{1}(1600)$ may produce an angular asymmetry similar
to that of Section 5, although the appearance of third pion makes the analysis more complicated.

\vspace{0.5cm}
\section{Diffractive production of the hybrid at large energy}
\vspace{0.5cm}

\noindent
At large energy, one should consider a different framework, namely
the impact representation \cite{serbo}  where the meson electroproduction amplitude
is factorized in impact factors and a reggeized two or three gluon exchange
known as perturbative Pomeron or Odderon exchange. Without performing
a detailed phenomenology of this reaction in this regime let us recall some well
known formula and briefly propose a strategy to help future (or present)
experiments at high energy to search for hybrids.

\noindent
The even charge conjugation of the hybrid meson selects in this case 
the Odderon exchange \cite{Ewerz} and the amplitude for its electroproduction is equal to
\begin{equation}  \label{odd}
\mathcal{M}_O =-\frac{8\,\pi^2\,s}{3!}\int\,\frac{d^2{\bf k}_1 \, d^2
{\bf k}_2\, d^2{\bf k}_3\, \delta^{(2)}({\bf k}_1 +{\bf k}_2 +{\bf k}_3-{\bf p}%
_{H^0})}{(2\pi)^6\,{\bf k}_1^2\,{\bf k}_2^2\,{\bf k}_3^2} J_O^{\gamma^*
\rightarrow H^0}\cdot J_O^{N \rightarrow N^{\prime}}
\end{equation}
where $J_O^{\gamma^* \rightarrow H^0}({\bf k}_1,{\bf k}_2,{\bf k}_3)$
and $J_O^{N \rightarrow N^{\prime}}({\bf k}_1,{\bf k}_2,{\bf k}_2)$ are the
impact factors  for the transition $\gamma^* \to H^0$ via Odderon
exchange  and of the nucleon in initial state $N$ into the nucleon in the
final state $N^{\prime}$.
 The impact factors are defined as the
$s-$channel discontinuities of the corresponding $S-$matrices describing
the $\gamma^* O \to
H^0$ and $N O \to
N'$ processes projected on the longitudinal (nonsense) polarizations
of the virtual gluons in the $t-$channel.

\noindent
The upper impact factors are calculated by the use of standard methods. The leading order
calculation in pQCD  gives in the case of a longitudinal polarized photon :
\begin{equation}
J_{O}^{\gamma _{L}^{*}}({\bf k}_{1},{\bf k}_{2},{\bf k}_{3})=-\frac{%
i\,e\,g^{3}\,d^{abc}\,Q}{4\,N_{C}}\;\int\limits_{0}^{1}\,dy\,y{\bar{y}}%
\,P_{O}({\bf k}_{1},{\bf k}_{2},{\bf k}_{3})\,\frac{1}{3}\Phi ^{H^0}(y)  \label{IFOL}
\end{equation}
where ${\bf k}_{1}+{\bf k}_{2}+{\bf k}_{3}={\bf p}_{H^0}$ and
\begin{eqnarray}
&&P_{O}({\bf k}_{1},{\bf k}_{2},{\bf k}_{3})=\frac{1}{y^{2}{\bf p}_{H^0
}^{\;2}+\mu ^{2}}-\frac{1}{{\bar{y}}^{2}{\bf p}_{H^0 }^{\;2}+\mu ^{2}}
\nonumber \\
&&-\sum\limits_{i=1}^{3}\left( \frac{1}{({\bf k}_{i}-y{\bf p}_{H^0
})^{2}+\mu ^{2}}-\frac{1}{({\bf k}_{i}-{\bar{y}}{\bf p}_{H^0 })^{2}+\mu ^{2}
}\right)
\end{eqnarray}
The proton impact factor cannot be calculated within perturbation theory. One may
 use phenomenological eikonal models of these impact
factors proposed in Ref \cite{protonO} which read
\begin{equation}
\label{IFPO}
J_{O}^{N\rightarrow N^{\prime }}=-i\frac{{\bar{g}}^{3}\,d^{abc}}{4\,N_{C}}
\,3\left[ F({\bf p}_{H^0 },0,0)-\sum\limits_{i=1}^{3}F({\bf k}_{i},{\bf p}
_{H^0 }-{\bf k}_{i},0)+2\,F({\bf k}_{1},{\bf k}_{2},{\bf k}_{3})\right]
\end{equation}
where
\begin{equation}
F({\bf k}_{1},{\bf k}_{2},{\bf k}_{3})=\frac{A^{2}}{A^{2}+\frac{1}{2}\left[ (
{\bf k}_{1}-{\bf k}_{2})^{2}+({\bf k}_{2}-{\bf k}_{3})^{2}+({\bf k}_{3}-
{\bf k}_{1})^{2}\right] }
\end{equation}
and $A=\frac{m_{\rho }}{2}$.
In these equations we  denote the soft QCD-coupling constant by
${\bar g}$ and one may take ${\alpha}_{soft} =\bar{g}^2/(4\pi)=0.5$ as a reasonable mean value.

\noindent
Since the Odderon amplitude is known to be rather small, producing the hybrid in
electroproduction at large energy
will be rather difficult. It will thus be useful to search for the
hybrid meson in this context through an interference of a Pomeron mediated amplitude to
a Odderon mediated one, as discussed in {\cite{PO}}. The three pion channel discussed
in the preceeding section is interesting in this respect since a charge asymmetry between
the $\pi^+$¥ and the $\pi^-$ will single out this interference. The charge asymmetry to measure
is defined as a  integral weighted with an antisymmetric function in the exchange
$\zeta^+ \to \zeta^-$, the simplest example being  $\zeta^+ - \zeta^-$. :
\begin{equation}
A(Q^2, t, m_{3\pi}^2) =\frac{\int\limits_{0}^{1} \,d\zeta^+\,d\zeta^-
(\zeta^+-\zeta^-)\,2\ \mbox{Re}
\left[{\cal M}_P^{\gamma^*_L}({\cal M}_O^{\gamma^*_L})^*  \right]}{
\int\limits_{0}^{1}\,d\zeta^+\,d\zeta^- \left[ |{\cal
M}_P^{\gamma^*_L}|^2
+ |{\cal M}_O^{\gamma^*_L}|^2  \right]     }
\end{equation}
which  is approximately equal to the ratio of the Odderon exchange amplitude and the Pomeron
exchange amplitude, which one may approximate to the production of a $J^{PC} = 1^{--}$¥ state
like $\omega(1650)$.

\noindent
This may be related to a forward-backward asymmetry in the rest frame of the two
charged pions, the forward direction being defined as the direction of the neutral $\pi^0$
meson. Defining the $\theta $ angle as the angle of the $\pi^+$ to the $\pi^0$ 3-momenta
in this frame, one gets
\begin{eqnarray}
A(Q^2, t, m_{3\pi}^2)
=\frac{ \int \cos \theta \,d \sigma
(s,Q^2,t,m_{3\pi}^2,\theta)}{
\int d \sigma (s,Q^2,t,m_{3\pi}^2,\theta)}
=\frac{2 \int\limits_{-1}^{1}\,d(\cos \theta) \,\cos \theta\, \Re
\left[{\cal M}_P^{\gamma^*_L}({\cal M}_O^{\gamma^*_L})^*  \right]}{
\int\limits_{-1}^{1}\,d\cos \theta \left[ |{\cal M}_P^{\gamma^*_L}|^2
+ |{\cal M}_O^{\gamma^*_L}|^2  \right]}
\end{eqnarray}

\vspace{.5cm}
\section{Conclusion}
\vspace{.5cm}

\noindent
In conclusion, we have calculated in this paper
the leading twist contribution to exotic hybrid meson with $J^{PC}=1^{-+}$ electroproduction
amplitude in the deep exclusive region.  The resulting order of magnitude
is somewhat smaller than the $\rho$ electroproduction but
similar to the $\pi$ electroproduction. The obtained cross section is
sizeable and should be measurable at dedicated experiments at JLab, Hermes or Compass.

\noindent
We made a systematic comparison with the non-exotic vector meson production.
To take into account NLO corrections, the differential cross-sections for these processes have
been computed using the BLM prescription for the renormalization scale.
In the case of $\rho$ production, our estimate is not far from a previous one  which
took into account kinematical higher twist corrections.

\noindent
We have also discussed in detail the $\pi\eta$ mode corresponding to the $\pi _{1}(1400)$
candidate in the reaction $e\, p\to e\,p\,\pi^0\eta$.
We have calculated an angular asymmetry  implied by charge conjugation properties
and got a sizeable hybrid effect which may be experimentally checked.

\noindent
In the region of  small $Q^2$  higher twist contributions should be carefully studied and
included. Note that they have already been considered in the case of deeply
virtual Compton scattering \cite{APT} where their presence was dictated by gauge invariance, and
for transversely polarized vector mesons \cite{AT} where the leading twist component vanishes.
We  leave this study for future works.

\noindent
Finally, the diffractive production at very high energy has been briefly studied. The weakness
of Odderon mediated processes makes the study of hybrid meson electroproduction a very difficult
task for HERA experiments.

\vspace{.5cm}
\section{ Acknowledgments}
\vspace{.5cm}

\noindent We acknowledge useful discussions with A.~Bakulev, I.~Balitsky, V.~Braun, M.~Diehl,
G.~Korchemsky, C.~Michael, S.~Mikhailov and O.~Pene. 
This work is supported in part by INTAS (Project 00/587)
and RFBR (Grant 03-02-16816).
The work of B.~P., L.~Sz. and S. W. is partially supported by the French-Polish scientific 
agreement Polonium and the Joint Research Activity "Generalised Parton 
Distributions" of the european I3 program Hadronic Physics, contract 
RII3-CT-2004-506078. 
I.~V.~A. thanks  NATO for a Grant. L.~Sz. thanks CNRS for a Grant supported 
his visit to LPT in Orsay.
L.~Sz. is a Visiting Fellow of the Fonds National pour la Recherche
Scientifique (Belgium)

\vspace{.5cm}
\section*{Appendix A: Functions $B_{11}(m_{\pi\eta}^2)$ and $B_{12}(m_{\pi\eta}^2)$}
\vspace{.5cm}

\noindent
We  now proceed to the calculation of the functions
$B_{11}(m_{\pi\eta}^2)$ and $B_{12}(m_{\pi\eta}^2)$   related to the corresponding
Breit-Wigner amplitudes.
Let us start from the consideration of  the
$\pi\eta$--to--vacuum matrix element of some vector nonlocal quark operator. We have
\begin{eqnarray}
\label{me1}
\langle \pi(p_\pi)\,\eta(p_\eta)| {\cal O}^V_{\mu}(-z;z)|0\rangle,
\end{eqnarray}
where ${\cal O}^V_\mu=\bar\psi(-z/2)\gamma_\mu\psi(z/2)$. This matrix element can be rewritten
in the equivalent form:
\begin{eqnarray}
\label{me2}
\langle \pi(p_\pi)\,\eta(p_\eta)| H(p)\rangle
\frac{1}{M^2_H-p^2-i\Gamma_H M_H}
\langle H(p)| {\cal O}^V_{\mu}(-z;z)|0\rangle +(\,other\,\, reson.\,).
\end{eqnarray}
We will, from now on, neglect the contribution from other resonances.
Note that owing to the momentum conservation law we have
$p^2=(p_\pi+p_\eta)^2=m^2_{\pi\eta}$.

\noindent
Further, we introduce the parameterization of the relevant matrix elements:
\begin{eqnarray}
\label{par}
&&\langle \pi(p_\pi)\,\eta(p_\eta)| H(p)\rangle=
G^{(-)}_{H\pi\eta}(m^2_\pi, m^2_\eta, M^2_H)(p_\pi-p_\eta)\cdot e^{(\lambda)}=
-ig_{H\pi\eta}(p_\pi-p_\eta)\cdot e^{(\lambda)},
\nonumber\\
&&\langle H(p)| {\cal O}^V_{\mu}(-z;z)|0\rangle=
if_H M_H\frac{e^{*\,(\lambda)}\cdot z}{p\cdot z}p_{\mu}
\int\limits_{0}^{1}dy\,e^{i(1-2y)p\cdot z/2}\phi_L^H(y),
\nonumber\\
&&\langle \pi(p_\pi)\,\eta(p_\eta)| {\cal O}^V_{\mu}(-z;z)|0\rangle=(p_\pi+p_\eta)_{\mu}
\int\limits_{0}^{1}dy\,
e^{i(1-2y)(p_\pi+p_\eta)\cdot z/2}\Phi^{(\pi\eta)}(y,\tilde\zeta, m_{\pi\eta}^2).
\end{eqnarray}
Using the above-mentioned parameterization,  equations (\ref{me1}) and (\ref{me2})
may be rewritten as
\begin{eqnarray}
\label{eq}
&&(p_\pi+p_\eta)_{\mu}
\int\limits_{0}^{1}dy\,e^{i(1-2y)(p_\pi+p_\eta)\cdot z/2}
\Phi^{(\pi\eta)}(y,\tilde\zeta, m_{\pi\eta}^2)=
\\
&&if_H M_H\, (-i)g_{H\pi\eta}\,
p_{\mu}\, \frac{z^\alpha}{p\cdot z}(p_\pi-p_\eta)^\beta\,
\frac{\sum\limits_\lambda e^{*\,(\lambda)}_\alpha e^{(\lambda)}_\beta}
{M^2_H-m^2_{\pi\eta}-i\Gamma_H M_H}\,
\int\limits_{0}^{1}dy\,e^{i(1-2y)p\cdot z/2}\phi_L^H(y).
\nonumber
\end{eqnarray}
The summation over polarization vectors reads
\begin{eqnarray}
\label{sumpol}
\sum\limits_{\lambda}e^{*\,(\lambda)}_\alpha e^{(\lambda)}_\beta=-g_{\alpha\beta}+
\frac{p_\alpha\,p_\beta}{M^2_H},
\end{eqnarray}
therefore the contraction of corresponding vectors with (\ref{sumpol}) gives us
\begin{eqnarray}
\frac{z^\alpha}{p\cdot z}(p_\pi-p_\eta)^\beta \biggl(
-g_{\alpha\beta}+
\frac{p_\alpha\,p_\beta}{M^2_H}\biggr)=
-\frac{(p_\pi-p_\eta)^+}{(p_\pi+p_\eta)^+}+\frac{m^2_\pi-m^2_\eta}{M^2_H}.
\end{eqnarray}

\noindent
Further, the term $B_{11}$ corresponding to the hybrid meson, see (\ref{approx}),
can be rewritten in the form:
\begin{eqnarray}
\Phi^{(\pi\eta)}_{1^{-+}}(y,\tilde\zeta,m^2_{\pi\eta})=18y(1-y)(2y-1)
B_{11}(m^2_{\pi\eta})P_1(\cos\theta),
\end{eqnarray}
where eqn. (\ref{2zeta}) is used.
Inserting this function into the {\it lhs} of (\ref{eq}) we get an explicit expression for the function
$B_{11}$:
\begin{eqnarray}
\label{B11}
&&B_{11}(m^2_{\pi\eta})=- \frac{{\cal F}}{18\,{\cal G}}\,
\frac{g_{H\pi\eta}f_H M_H}{M^2_H-m^2_{\pi\eta}-i\Gamma_H M_H}\,
\frac{\lambda(m^2_{\pi\eta}, m^2_\eta, m^2_\pi)}{m^2_{\pi\eta}}
\nonumber\\
&&\Biggl[
\biggl(
1-\frac{m^2_\pi-m^2_\eta}{m^2_{\pi\eta}}\,
\frac{(p_\pi+p_\eta)^+}{(p_\pi-p_\eta)^+}
\biggr)^{-1} -
\frac{m^2_\pi-m^2_\eta}{M^2_H}
\biggl(
\frac{(p_\pi-p_\eta)^+}{(p_\pi+p_\eta)^+} -
\frac{m^2_\pi-m^2_\eta}{m^2_{\pi\eta}}
\biggr)^{-1}
\Biggr],
\end{eqnarray}
where
\begin{eqnarray}
&&{\cal F}=\int\limits_{0}^{1}dy\,e^{i(1-2y)p\cdot z/2}\,\phi_L^H(y),
\nonumber\\
&&{\cal G}=\int\limits_{0}^{1}dy\,e^{i(1-2y)(p_\pi+p_\eta)\cdot z/2}y(1-y)(2y-1).
\end{eqnarray}

\noindent
If we use the asymptotic form for the function $\phi_L^H(y)$,
the expression (\ref{B11}) is rewritten as
\begin{eqnarray}
\label{B11W}
&&B_{11}(m^2_{\pi\eta})=\frac{5}{3}\,
\frac{g_{H\pi\eta}f_H M_H}{M^2_H-m^2_{\pi\eta}-i\Gamma_H M_H}\,
\frac{\lambda(m^2_{\pi\eta}, m^2_\eta, m^2_\pi)}{m^2_{\pi\eta}}
\nonumber\\
&&\Biggl[
\biggl(
1-\frac{m^2_\pi-m^2_\eta}{m^2_{\pi\eta}}\,
\frac{(p_\pi+p_\eta)^+}{(p_\pi-p_\eta)^+}
\biggr)^{-1} -
\frac{m^2_\pi-m^2_\eta}{M^2_H}
\biggl(
\frac{(p_\pi-p_\eta)^+}{(p_\pi+p_\eta)^+} -
\frac{m^2_\pi-m^2_\eta}{m^2_{\pi\eta}}
\biggr)^{-1}
\Biggr],
\end{eqnarray}
where the value $m^2_{\pi\eta}$ is in the vicinity of the hybrid mass $M_H$.
We thus have obtained the following expression for the function $B_{11}$
\begin{eqnarray}
\label{B11F}
B_{11}(m^2_{\pi\eta})\biggl|_{m^2_{\pi\eta}\approx M^2_H}=
\frac{5}{3}\,
\frac{g_{H\pi\eta}f_H M_H}{M^2_H-m^2_{\pi\eta}-i\Gamma_H M_H}
\frac{\lambda(m^2_{\pi\eta}, m^2_\eta, m^2_\pi)}{m^2_{\pi\eta}}
\biggl|_{m^2_{\pi\eta}\approx M^2_H}.
\end{eqnarray}

\noindent
We will now focus on the calculation of the function $B_{12}(m^2_{\pi\eta})$.
As mentioned above, the $a_2$-resonance formula for the case
$m^2_{\pi\eta}\approx M^2_{a_2}$ reads (cf. (\ref{me1}) and (\ref{me2}))
\begin{eqnarray}
\label{mea2}
&&\langle \pi(p_\pi)\,\eta(p_\eta)| {\cal O}^V_{\mu}(-z;z)|0\rangle=
\\
&&\langle \pi(p_\pi)\,\eta(p_\eta)| a_2(p)\rangle
\frac{1}{M^2_{a_2}-p^2-i\Gamma_{a_2} M_{a_2}}
\langle a_2(p)| {\cal O}^V_{\mu}(-z;z)|0\rangle.
\nonumber
\end{eqnarray}
The parameterizations of matrix elements standing in (\ref{mea2})
can be introduced in the following forms. First of all,
we write the parameterization for the $\pi\eta$--to--$a_2$ matrix element:
\begin{eqnarray}
\label{para21}
&&\langle \pi(p_\pi)\,\eta(p_\eta)| a_2(p)\rangle=
e^{(\lambda)}_{\mu\nu} {\cal V}^{\mu\nu},
\end{eqnarray}
where
\begin{eqnarray}
e^{(\lambda)}_{\mu\nu}p^{\mu}=e^{(\lambda)}_{\mu\nu}p^{\nu}=0,\quad
e^{(\lambda)}_{\mu\nu}g^{\mu\nu}=0,\quad e^{(\lambda)}_{\mu\nu}=e^{(\lambda)}_{\nu\mu}.
\end{eqnarray}
Due to the Lorentz invariance the most general representation of
the tensor ${\cal V}^{\mu\nu}$ take the form
\begin{eqnarray}
{\cal V}^{\mu\nu}&=&
G^{(1)}_{a_2\pi\eta}(m^2_\pi, m^2_\eta, M^2_{a_2})p_\pi^\mu p_\pi^\nu+
G^{(2)}_{a_2\pi\eta}(m^2_\pi, m^2_\eta, M^2_{a_2})p_\pi^\mu p_\eta^\nu+
\nonumber\\
&&G^{(3)}_{a_2\pi\eta}(m^2_\pi, m^2_\eta, M^2_{a_2})p_\eta^\mu p_\pi^\nu+
G^{(4)}_{a_2\pi\eta}(m^2_\pi, m^2_\eta, M^2_{a_2})p_\eta^\mu p_\eta^\nu.
\end{eqnarray}
We fix the form factors $G^{(i)}$ as
\begin{eqnarray}
&&G^{(1)}_{a_2\pi\eta}(m^2_\pi, m^2_\eta, M^2_{a_2})=
G^{(4)}_{a_2\pi\eta}(m^2_\pi, m^2_\eta, M^2_{a_2})=i g_{a_2\pi\eta},
\nonumber\\
&&G^{(2)}_{a_2\pi\eta}(m^2_\pi, m^2_\eta, M^2_{a_2})=
G^{(3)}_{a_2\pi\eta}(m^2_\pi, m^2_\eta, M^2_{a_2})=-i g_{a_2\pi\eta}.
\end{eqnarray}
In this case the parameterization of (\ref{para21}) is reduced to the form
\begin{eqnarray}
\label{para22}
&&\langle \pi(p_\pi)\,\eta(p_\eta)| a_2(p)\rangle=
i g_{a_2\pi\eta} e^{(\lambda)}_{\mu\nu} \,(p_\pi-p_\eta)^\mu\,(p_\pi-p_\eta)^\nu.
\end{eqnarray}
The parameterization of vacuum--$a_2$-meson matrix element can be written as
\begin{eqnarray}
\label{para2}
\langle a_2(p)| {\cal O}^V_{\mu}(-z;z)|0\rangle=
f_{a_2} M_{a_2}^2\frac{e^{*\,(\lambda)}_{\alpha\beta}\,z^{\alpha}z^{\beta}}{(p\cdot z)^2}p_{\mu}
\int\limits_{0}^{1}dy\,e^{i(1-2y)p\cdot z/2}\phi_L^{a_2}(y).
\end{eqnarray}
Again, the asymptotic form of function $\phi_L^{a_2}(y)$ can be defined as
\begin{eqnarray}
\label{asa2}
\Phi_L^{a_2}(y)=30y(1-y)(2y-1).
\end{eqnarray}

\noindent
Further, the $a_2$-resonance part of $\pi\eta$ distribution amplitude reads
\begin{eqnarray}
\label{B12p}
\Phi^{(\pi\eta)}_{2^{++}}(y,\tilde\zeta,m^2_{\pi\eta})=18y(1-y)(2y-1)
B_{12}(m^2_{\pi\eta})P_2(\cos\theta).
\end{eqnarray}

\noindent
Thereafter,
inserting the parametrical representations of hadron matrix elements
(\ref{para22}), (\ref{para2}) in  eqn. (\ref{mea2}) with (\ref{asa2}), (\ref{B12p}),
one can see that
\begin{eqnarray}
&&B_{12}(m^2_{\pi\eta}) P_2(\cos\theta)=
\\
&&\frac{30}{18} \frac{i  g_{a_2\pi\eta} f_{a_2} M^2_{a_2}}
{M^2_{a_2}-m^2_{\pi\eta}-i\Gamma_{a_2} M_{a_2}}\,
\biggl[
(p_\pi-p_\eta)^\mu\,(p_\pi-p_\eta)^\nu
\frac{z^\alpha z^\beta}{(p\cdot z)^2}\,
\sum\limits_{\lambda} e^{(\lambda)}_{\mu\nu} e^{*\,(\lambda)}_{\alpha\beta}
\biggr],
\nonumber
\end{eqnarray}
where as usual
\begin{eqnarray}
\label{pol}
&&\sum\limits_{\lambda}e^{(\lambda)}_{\mu\nu}e^{*\,(\lambda)}_{\alpha\beta}=
\frac{1}{2}X_{\mu\alpha}X_{\nu\beta}+\frac{1}{2}X_{\mu\beta}X_{\nu\alpha}-
\frac{1}{3}X_{\mu\nu}X_{\alpha\beta},
\nonumber\\
&& X_{\alpha_1\alpha_2}=-g_{\alpha_1\alpha_2}+\frac{p_{\alpha_1}p_{\alpha_2}}{M^2_{a_2}}.
\end{eqnarray}

\noindent
Using (\ref{2zeta}), the Legendre polynomial $P_2(\cos\theta)$
can be presented as
\begin{eqnarray}
&&P_2(\cos\theta)=\frac{3}{2}\cos^2\theta-\frac{1}{2}=
\\
&&\frac{m^4_{\pi\eta}}{2 \lambda^2(m^2_{\pi\eta}, m^2_\eta, m^2_\pi)}
\biggl(
3 (2\tilde\zeta-1)^2-\frac{\lambda^2(m^2_{\pi\eta}, m^2_\eta, m^2_\pi)}
{m^4_{\pi\eta}}
\biggr).
\nonumber
\end{eqnarray}
A straightforward computation leads to
\begin{eqnarray}
&&(p_\pi-p_\eta)^\mu\,(p_\pi-p_\eta)^\nu
\frac{z^\alpha z^\beta}{(p.z)^2}\,
\sum\limits_{\lambda} e^{(\lambda)}_{\mu\nu} e^{*\,(\lambda)}_{\alpha\beta}=
\frac{1}{3}\,\frac{2 P_2(\cos\theta) \lambda^2(m^2_{\pi\eta}, m^2_\eta, m^2_\pi)}{m^4_{\pi\eta}}=
\nonumber\\
&&\frac{1}{3}\,\biggl(
3 (2\tilde\zeta-1)^2-\frac{\lambda^2(m^2_{\pi\eta}, m^2_\eta, m^2_\pi)}
{m^4_{\pi\eta}}
\biggr),
\end{eqnarray}
where we used that
\begin{eqnarray}
z^2=0,  \quad
p_\pi\cdot p_\eta=\frac{M^2_{a_2}-m^2_\pi-m^2_\eta}{2}.
\end{eqnarray}
We have finally obtained the following representation for the function
$B_{12}(m^2_{\pi\eta})$ in the vicinity of $m^2_{\pi\eta}\approx M^2_{a_2}$:
\begin{eqnarray}
B_{12}(m^2_{\pi\eta})\biggr|_{m^2_{\pi\eta}\approx M^2_{a_2}}=
\frac{10}{9} \frac{i  g_{a_2\pi\eta} f_{a_2} M^2_{a_2}}
{M^2_{a_2}-m^2_{\pi\eta}-i\Gamma_{a_2} M_{a_2}}\,
\frac{\lambda^2(m^2_{\pi\eta}, m^2_\eta, m^2_\pi)}{m^4_{\pi\eta}}
\biggr|_{m^2_{\pi\eta}\approx M^2_{a_2}}.
\end{eqnarray}

\vspace{1.cm}

\end{document}